\documentclass[aps,prl,onecolumn,10pt,noshowpacs,noshowkeys, notitlepage,twoside,a4paper,nobibnotes,superscriptaddress]{revtex4-1}
\usepackage{color}
\usepackage[utf8]{inputenc}
\usepackage{amsmath}
\usepackage{amsfonts}
\usepackage{amssymb}
\usepackage{graphicx}
\usepackage{grffile}

\newcommand{\symkowm}[2]{ \left<\left[ #1 ; #2 \right]_{+} \right>}
\begin{document}
\title{Partition of energy for a dissipative quantum oscillator}
\author{P. Bialas}
\affiliation{Institute of Physics and Silesian Center for Education and Interdisciplinary Research, University of Silesia, 41-500 Chorz{\'o}w, Poland}
\author{J. Spiechowicz}
\affiliation{Institute of Physics and Silesian Center for Education and Interdisciplinary Research, University of Silesia, 41-500 Chorz{\'o}w, Poland}
\affiliation{Institute of Physics, University of Augsburg, D-86135 Augsburg, Germany}
\author{J. {\L}uczka}
\email[Correspondence and requests for materials should be addressed to J.{\L}.]{ (e-mail: jerzy.luczka@us.edu.pl)}
\affiliation{Institute of Physics and Silesian Center for Education and Interdisciplinary Research, University of Silesia, 41-500 Chorz{\'o}w, Poland}
\email{jerzy.luczka@us.edu.pl} 
\begin{abstract}
We reveal a new face of the old clich{\'e}d system: a dissipative quantum harmonic oscillator. 
We formulate and study a quantum counterpart of the energy equipartition theorem satisfied for classical systems.
Both mean kinetic energy $E_k$ and mean potential energy $E_p$ of the oscillator are expressed as 
$E_k = \langle \mathcal E_k \rangle$  and $E_p = \langle \mathcal E_p \rangle$, where 
$\langle \mathcal E_k \rangle$ and $ \langle \mathcal E_p \rangle$ are mean  kinetic and potential energies  per one degree of freedom of the thermostat which  consists of harmonic oscillators too. The symbol $\langle ...\rangle$ denotes two-fold averaging: 
 (i) over the Gibbs canonical state  for the thermostat  and (ii) over thermostat oscillators  frequencies $\omega$ which contribute to $E_k$ and $E_p$ according to the probability distribution $\mathbb{P}_k(\omega)$ and $\mathbb{P}_p(\omega)$, respectively.
The role of the system-thermostat coupling strength and the memory time is analysed for the exponentially decaying memory function (Drude dissipation mechanism) and the algebraically decaying damping kernel.
\end{abstract}
\maketitle

In classical physics a harmonic oscillator describes small oscillations. Its quantum version is a standard model to introduce  creation and annihilation Bose operators. In the theory of open quantum systems the harmonic oscillator is one of the simplest systems to investigate dissipation processes (see e.g. \cite{weis} and refs. therein) and decoherence phenomena (see e.g. \cite{zurek,schlosshauer} and refs. therein). It has attracted considerable interest over the last fifty years. 
Nevertheless there is still a plenty of room which is {\it terra incognita}. As an example, it has been lately applied in the problem of quantum-to-classical transition, formation of dynamical spectrum broadcast structures and classical objectivity as a property of quantum states \cite{korbicz}. Finally, we subjectively cite only a few papers \cite{boy,smirne,carlesso,china,lampo} published in the last two years to confirm that it is still the topic of active research. We also wish to revisit the dissipative quantum oscillator and discuss a quite different aspect, namely, the quantum counterpart of the theorem of energy equipartition (TEE) in classical statistical physics. Surely, it belongs to one of the fundamental laws which represents a universal relation in the sense that it does not depend on a number of particles in the system, a potential force which acts on them, any interaction between particles or the strength of coupling between the system and thermostat \cite{huang,terlecki}.  Beginning of its formulations is dated back to 19th century, to the times of James Clerk Maxwell and Ludwig Boltzmann. The latter in 1876 showed that average kinetic energy is equally shared in a portion of $E_k = k_B T/2$ among all degrees of freedom of a system \cite{boltzmann}.  Since that time the TEE has become one of the most important and most useful relation exploited in various branches of Natural Science. In contrast, this law is no longer true for quantum systems. 
From the time of Max Planck and birth of quantum physics  a 
quantum counterpart of TEE has not been explicitly proposed. Partial results on mean energy of some particular systems can be found in literature. Lately, we have derived an appealing formula which is a generalization of the classical TEE \cite{arxiv} 
In this case the mean kinetic energy is not shared equally among all accessible degrees of freedom. 
In contrast, the kinetic energy $E_k$ of a quantum harmonic oscillator is a thermally averaged kinetic energy per one degree of freedom of the thermostat oscillators. 
The latter contribute to $E_k$ with different degrees: if the thermostat oscillator has eigenfrequency $\omega$ then its input to $E_k$ is determined by the probability density  $\mathbb{P}_k(\omega)$. 
We study the impact of two dissipation mechanism determined by the exponentially and algebraically decaying dissipation function on properties of the probability distribution $\mathbb{P}_k(\omega)$ and the mean kinetic energy of the quantum harmonic oscillator. Similar analysis is performed for the mean potential energy of the system.
\section{Model and Results} 
We study the celebrated model of a quantum open system $S$, i.e. a quantum harmonic oscillator of mass $M$ and eigenfrequency $\omega_0$. It is in contact with a heat bath $B$ modelled as a collection of  independent quantum harmonic oscillators which form thermostat  of temperature $T$ being in an equilibrium Gibbs canonical state. The  Hamiltonian of such a composite  system  $S+B$ has the form \cite{weis,maga,ulersma,ingold,hujcen,chaos,ph,caldeira,hu} 
(a complete list of papers concerning this problem is too long and our choice is selective)  
\begin{equation}
H=\frac{p^2}{2M}+\frac{1}{2} M \omega_0^2 x^2+ \sum_i \left[ \frac{p_i^2}{2m_i} + \frac{m_i\omega_i^2}{2} \left( q_i - \frac{c_i x}{m_i \omega_i^2} \right)^2 \right],
\end{equation}
where the coordinate and momentum operators $\{x, p\}$ refer to the Brownian particle and $\{q_i, p_i\}$ are the coordinate and momentum operators of the $i$-th heat bath oscillator of mass $m_i$ and the eigenfrequency $\omega_i$. The parameter $c_i$ characterizes the coupling  strength of the central system $S$ with the $i$-th thermostat oscillator. 
All coordinate and momentum operators obey canonical equal-time commutation relations. From the Heisenberg equations of motion for all coordinate and momentum operators one can obtain an effective equation of motion for the oscillator coordinate operator $x(t)$.  It is called a generalized quantum Langevin equation (GQLE) and  reads \cite{bialas}
\begin{equation}\label{GLE}
M{\ddot x}(t)+ M \omega_0^2 x(t)+ \int_0^t du \; \gamma(t-u) \dot{x}(u) = -\gamma(t) x(0)+ \eta(t), 
\end{equation}
where dot denotes time derivative, $\gamma(t)$ is a dissipation function (damping or memory kernel) and $\eta(t)$ is  quantum noise,  
\begin{eqnarray} 
\gamma(t) &=& \int_0^{\infty} d \omega J(\omega) \cos(\omega t),   \label{g}\\
\eta(t) &=& \sum_i c_i \left[q_i(0) \cos(\omega_i t) + \frac{p_i(0)}{m_i \omega_i}
\sin(\omega_i t) \right]
\label{force} 
\end{eqnarray}  
and $J(\omega)$ is a spectral density of thermostat modes which contains all information on the system-thermostat coupling:
\begin{equation}\label{spectral}
J(\omega) = \sum_i \frac{c_i^2}{m_i \omega_i^2} \delta(\omega -\omega_i). 
\end{equation}
In the standard approach it is assumed that the initial state $\rho(0)$ of the composite  system $S+B$ is uncorrelated, i.e., \mbox{$\rho(0)=\rho_S(0)\,\otimes\,\rho_T(0)$}, where $\rho_S$ is an arbitrary state of the Brownian particle and $\rho_T$ is an equilibrium Gibbs canonical state of thermostat of temperature $T$. Next, the thermodynamic limit is imposed meaning that the thermal reservoir is infinitely extended. Then the singular spectral function $J(\omega)$ in Eq. (\ref{spectral}) tends to a (piecewise) continuous function.

Solving Eq. (\ref{GLE}) for $x(t)$ we can obtain the oscillator momentum operator $p(t)$ from the standard relation  \mbox{$p(t) = M \dot x(t)$.} It allows to calculate the mean kinetic $E_k(t) =  \langle p^2(t)\rangle/2M$ and potential $E_p(t) = M\omega_0^2 \langle x^2(t)\rangle/2$ energy of the quantum oscillator. In the long time limit, when the equilibrium state is reached, one gets the following expressions for the above quantities (see the section Methods)
\begin{eqnarray}
E_k &=& \lim_{t\to\infty} \frac{1}{2M} \langle p^2(t)\rangle = \langle \mathcal{E}_k \rangle = \int_0^{\infty} d\omega \; \mathcal{E}_k(\omega)\mathbb{P}_k(\omega), \label{Ek}\\
E_p &=& \lim_{t\to\infty} \frac{1}{2} M\omega_0^2 \langle x^2(t)\rangle =\langle \mathcal{E}_p \rangle = \int_0^{\infty} d\omega \; \mathcal{E}_p(\omega)\mathbb{P}_p(\omega), \label{Ep}
\end{eqnarray} 
where 
\begin{equation}\label{Ekp}
\mathcal{E}_k(\omega) = \mathcal{E}_p(\omega) = \frac{\hbar \omega}{4} \coth\left({\frac{\hbar \omega}{ 2k_BT}}\right) 
\end{equation} 
are thermally averaged kinetic and potential energies of one degree of freedom of the thermostat \cite{feynman}. The latter average is over the Gibbs canonical ensemble with the statistical operator $\rho_T \propto \mbox{exp}[-H_B/k_BT]$, where $H_B$ is the Hamiltonian of the heat bath  and $k_B$ is the Boltzmann constant. The probability distributions $\mathbb{P}_k(\omega)$ and $\mathbb{P}_p(\omega)$ have the form  
\begin{eqnarray}\label{P}
\mathbb{P}_k(\omega) = \frac{1}{\pi} \left[\hat{R}_L(i\omega) + \hat{R}_L(-i\omega) \right], \quad \quad 
\mathbb{P}_p(\omega) = \frac{i M \omega_0^2}{ \pi \omega} \left[\hat{Q}_L (i
\omega) - \hat{Q}_L (-i \omega) \right]
\end{eqnarray}
and  
\begin{eqnarray}\label{RL} 
\hat{R}_L(z) = \frac{z M}{Mz^2 + z \hat{\gamma}_L(z) + M \omega_0^2}, 
\quad \quad \hat{Q}_L(z) =  \frac{1}{Mz^2 + z \hat{\gamma}_L(z) + M \omega_0^2} 
\end{eqnarray} 
are Laplace transforms of the response functions $R(t)$ and $Q(t)$ for the momentum and coordinate operator of the oscillator, respectively. The function $\hat \gamma_L(z)$ is the Laplace transform of the damping kernel $\gamma(t)$. To be more precise, for any function $f(t)$ its Laplace transform is defined as 
\begin{equation}\label{fL} 
\hat f_L(z) = \int_0^{\infty} dt \; {\mbox e}^{-zt} f(t). 
\end{equation}  
Eqs. (\ref{Ek}) and (\ref{Ep}) are quantum counterparts of the theorem on the energy equipartition of classical systems. One can note that for quantum systems there is no equipartition but there is another form of partition of energy described by the corresponding frequency probability distributions:\\
(i) The mean kinetic energy $E_k$ of the quantum oscillator is a thermally averaged kinetic energy per one degree of freedom of the thermostat oscillators.\\
(ii) The mean potential energy $E_p$ of the quantum oscillator is a thermally averaged potential  energy per one degree of freedom of the thermostat oscillators.\\ \noindent
This should be contrasted  with the corresponding classical system for which   average  energy is equally shared in the same portion $k_B T/2$ among all degrees of freedom of the composite system, i.e., 
\begin{equation}
E_k =  \mathcal{E}_k = E_p = \mathcal{E}_p = k_B T/2. 
\end{equation}
According to our above statement in the quantum case the kinetic energy is not divided equally among all degrees of freedom and thermostat oscillators contribute to $E_k$ with a different degree, i.e. if the thermostat oscillator has eigenfrequency $\omega$ then its contribution to $E_k$ is determined by the probability density $\mathbb{P}_k(\omega)$. 
Because the model is exactly solvable the probability density  $\mathbb{P}_k(\omega)$ is exact and determined by the Laplace transform $\hat{R}_L(z)$ of the response function $R(t)$. It contains the Laplace transform $\hat{\gamma}_L(z)$ of the memory function $\gamma(t)$ which, via Eq. (\ref{g}), depends on the spectral function $J(\omega)$  which in turn, via Eq. (\ref{spectral}), comprises  all information on the oscillator-thermostat interaction and frequencies of the bath modes. This argumentation applies, {\it mutatis mutandis}, to the mean potential energy of the oscillator.

We now consider two random variables $\xi_k$ and $\xi_p$ distributed according to the probability density $\mathbb{P}_k(\omega)$ and  $\mathbb{P}_p(\omega)$, respectively.  
The first moments, i.e. the mean values $\langle \xi_k \rangle$ and $\langle \xi_p \rangle$ of the random variables $\xi_k$ and $\xi_p$ are proportional to the kinetic $E_k$ and potential  $E_p$ energy of the oscillator at zero temperature 
$T = 0$, namely
\begin{equation}
	E_k^0 = E_k(T = 0) = \frac{\hbar}{4}  \, \langle \xi_k  \rangle \quad \quad  
	E_p^0 = E_p(T = 0) = \frac{\hbar}{4}  \, \langle \xi_p \rangle. 
\end{equation}
Although absolute temperature of the environment $B$ is zero the central system $S$ is strongly influenced by the purely quantum vacuum fluctuations of the bath and therefore its energy is always greater than zero. Hereafter, we analyse the influence of dissipation mechanisms modelled by two memory kernels, the exponentially decaying function $\gamma_D(t)$ (the Drude model) and the algebraically decaying one $\gamma_A(t)$, namely,   
\begin{eqnarray}\label{gamma} 
\gamma_D(t) = \frac{\gamma_0}{2 \tau_c}e^{-t/\tau_c}, \quad \quad 
\gamma_A(t) = \frac{\gamma_0}{\pi}\, \frac{\tau_c}{t^2 + \tau_c^2}.
\end{eqnarray} 
%
The corresponding form of the spectral density $J(\omega)$ is  obtained from Eqs. (\ref{g})  and (\ref{Four}): 
\begin{eqnarray} \label{JDA}
J_D(\omega) = \frac{1}{\pi}\frac{\gamma_0}{1 + \omega^2\tau_c^2}, \quad \quad 
J_A(\omega) = \frac{\gamma_0}{\pi}e^{-\omega \tau_c}.
\end{eqnarray}
In the above scaling, if the memory time $\tau_c \to 0$,   both functions $\gamma_D(t)$ and $\gamma_A(t)$ tend to the Dirac delta and the integral term in the GQLE (\ref{GLE})  reduces to the frictional force of the Stokes form. 
For classical systems it corresponds to the limit of Gaussian white noise when thermal noise is $\delta$-correlated. There are four parameters: $M, \gamma_0, \tau_c$ and $\omega_0$ and three characteristic times (or frequencies being their reciprocals): \mbox{$\tau_v = M/\gamma_0, \tau_c,  1/\omega_0$}. If we rescale all quantities to the dimensionless form then there are only two dimensionless parameters 
\begin{equation} \label{alfa}
\alpha =  \frac{M}{\tau_c \gamma_0} =\frac{\tau_v}{\tau_c}, \quad \quad 
\tilde{\omega}_0 =\omega_0 \tau_v,
\end{equation}
where $\alpha$ is a ratio of two characteristic times $\tau_v$ and $\tau_c$. There is an alternative scaling with $\hat{\omega}_0 = \omega_0 \tau_c$ but since we will be interested mainly in the role of the memory time we use only (\ref{alfa}). We would like to pay attention that in this scaling the parameter $\tau_v$ is fixed and the change of $\alpha$ means the change of the memory time $\tau_c$. 

\begin{figure}[t]
	\centering
	\includegraphics[width=0.99\linewidth]{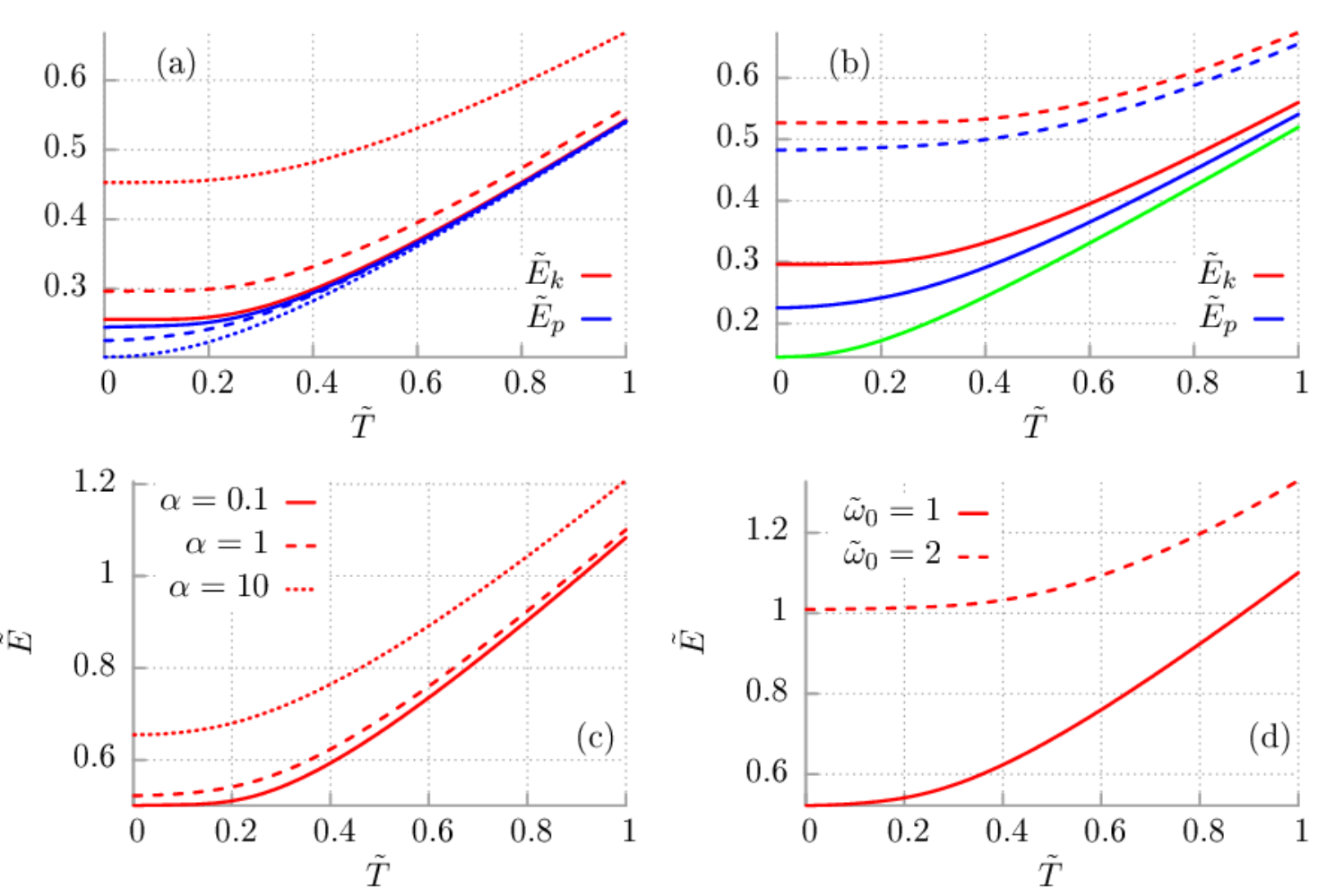}
		\caption{Drude model of dissipation.  The dimensionless mean kinetic energy $\tilde E_k= \tau_v {E}_k/\hbar$ (red) and  mean potential energy  $\tilde E_p= \tau_v {E}_p/\hbar$ (blue)  versus dimensionless temperature $\tilde T = \tau_v k_B T/\hbar$, 
	where $\tau_v=M/\gamma_0$ is fixed.  Panel (a): Solid line $\alpha = \tau_v/\tau_c = 0.1$, dashed line: $\alpha = 1$, dotted line $\alpha = 10$; all for the fixed eigenfrequency   $\tilde{\omega}_0 = \omega_0 \tau_v = 1$. Panel (b): Solid line $\tilde{\omega}_0 = 1$, dashed line $\tilde{\omega}_0 = 2$ and  fixed $\alpha = 1$. The exception here is the green solid line which shows the mean kinetic energy $\tilde{E}_k$ for the free Brownian particle with $\tilde{\omega}_0 = 0$.
Panel (c): The total energy  $\tilde E= \tilde E_k + \tilde E_p$ corresponding to the regime of panel (a). Panel (d):	The total energy corresponding to the regime of panel (b).}
	\label{fig1}
\end{figure}

\subsection{Mean kinetic and potential energy}

First, we consider the Drude model for which 
\begin{equation} \label{D_k}
\mathbb{P}_k(\omega) = \frac{2}{\pi} \frac{2 \gamma_0  M \omega^2}{ \omega^2 [\gamma_0 + 2 M \tau_c \left(\omega_0^2-\omega^2\right)]^2 +4 M^2 \left(\omega^2-\omega_0^2\right)^2 }.  
\end{equation}
%
%
%
In the case of algebraic decay of $\gamma(t)$ as it is in Eq. (\ref{gamma}), it takes the form 
\begin{equation}
    \label{C_k}
    \mathbb{P}_k(\omega) = \frac{4\pi \gamma_0 M \omega^2 e^{-\omega \tau_c}}{C_1(M, \gamma_0, \tau_c, \omega_0, \omega)C_2(M, \gamma_0, \tau_c, \omega_0, \omega)}
\end{equation}
with
\begin{eqnarray}
    C_1(M, \gamma_0, \tau_c, \omega_0, \omega) &=& 2\pi M (\omega^2 - \omega_0^2) + 2\gamma_0\,\omega\,\mbox{Ci}(i\omega \tau_c)\sinh{(\omega \tau_c)} \nonumber + \gamma_0\,\omega\cosh{(\omega \tau_c)}\left(-i\pi - 2\mbox{Shi}(\omega \tau_c)\right),\\
    C_2(M, \gamma_0, \tau_c, \omega_0, \omega) &=& 2\pi M (\omega^2 - \omega_0^2) + 2\gamma_0\, \omega\,\mbox{Ci}(-i \omega \tau_c)\sinh{(\omega \tau_c)} \nonumber +  i\gamma_0\, \omega \cosh{(\omega \tau_c)}\left(\pi + 2i\mbox{Shi}(\omega \tau_c)\right)
\end{eqnarray}
and  
\begin{equation}
\mbox{Ci}(z) = -\int_z^\infty dt\, \frac{\cos{t}}{t}, \quad \quad
\mbox{Shi}(z) = \int_0^z dt\, \frac{\sinh{t}}{t}.
\end{equation}
%
The expressions for the corresponding $\mathbb{P}_p(\omega)$ can be obtained from Eq. (\ref{D_k}) or (\ref{C_k}) by changing $\omega^2 \to \omega_0^2$ in their numerators. 
In all figures, we use dimensionless quantities and parameters. In particular, the rescaled probability densities 
$\mathbb{\tilde{P}}_k(x) = (1/\tau_v) \mathbb{P}_k(x/\tau_v)$ and 
$\mathbb{\tilde{P}}_p(x) = (1/\tau_v) \mathbb{P}_p(x/\tau_v)$,  where $x=\omega \tau_v$ is a dimensionless frequency and $\tau_v$ is fixed. In consequence, the change of  the parameter $\alpha=\tau_v/\tau_c$ denotes the change of the memory time $\tau_c$.

\begin{figure}[t]
	\centering
	\includegraphics[width=0.99\linewidth]{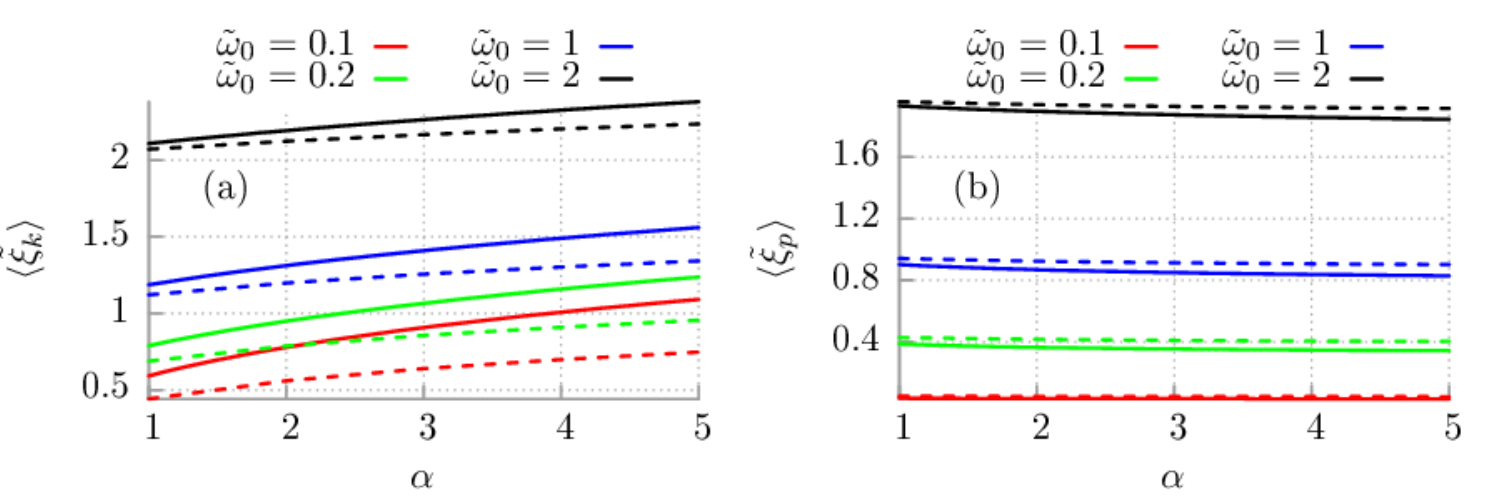}
		\caption{Panel (a): the mean value $\langle \tilde{\xi}_k \rangle = \tau_v \langle \xi_k \rangle$ of the random variable distributed according to the probability distribution $\mathbb{\tilde{P}}_k(x) = (1/\tau_v) \mathbb{P}_k(x/\tau_v)$ corresponding to the mean kinetic energy of the quantum harmonic oscillator is shown as the function of the parameter $\alpha = \tau_v/\tau_c$, where $\tau_v=M/\gamma_0$ is fixed,  and different values of eigenfrequency $\tilde{\omega}_0 = \omega_0 \tau_v$. Panel (b): the first statistical moment of the probability density $\mathbb{{\tilde P}}_p(x)= (1/\tau_v) \mathbb{P}_p(x/\tau_v)$ for the potential energy of the quantum harmonic oscillator. Solid lines correspond to the Drude (exponential) model and dashed lines to algebraic decay of $\gamma(t)$.}
	\label{fig2}
\end{figure} 

In Fig. 1 we illustrate the mean kinetic and potential energy  determined by Eq. (\ref{Ek}) and (\ref{Ep}), respectively, 
as a function of temperature for selected values of the model parameters. In particular, in panel (a) we present the influence of the memory time $\tau_c$ via the parameter $\alpha=\tau_v/\tau_c$ with fixed $\tau_v=M/\gamma_0$  and the oscillator eigenfrequency $\tilde{\omega}_0 = 1$. We note that regardless of the value of the memory time for this set of parameters the potential energy is always smaller than the kinetic one. Moreover, when the memory time decreases (i.e. $\alpha$ increases) the kinetic energy increases whereas the potential one is decreasing. On the other hand if time $\tau_c$ increases (i.e. $\alpha$ decreases) then the difference between the kinetic and potential energy is getting smaller and smaller and in the limit of infinitely long memory time it tends to zero. Alternatively, if the memory time $\tau_c$ is fixed and we change $\tau_v=M/\gamma_0$ in  $\alpha = \tau_v/\tau_c$ we observe that the kinetic and potential energy  approaches the same value  in the limit of large values of $\alpha$ (not depicted). It implies that either (i) the mass $M$ of the particle is large or (ii) the coupling $\gamma_0$ between the system and thermostat is weak. In the latter situation one could say that the system may be approximated by a free harmonic oscillator, which especially in the low temperature limit approaches a coherent state, where the position and momentum variances (proportional to kinetic and potential energy) match. The problem of relation between the kinetic and potential energy is discussed also in Ref. [19].

In  panel (b) of Fig. 1 we present the same characteristics but now depicted for the fixed memory time $\alpha = 1$ and different values of the oscillator eigenfrequency $\tilde{\omega}_0$. The observation is that for increasing values of the latter parameter both the kinetic and potential energy is growing. However, still the kinetic one is larger than the potential energy. The reader should note there also the interesting comparison with the case of a free quantum Brownian particle $\tilde{\omega}_0 = 0$ which is marked by the green colour. It turns out that the kinetic energy of a quantum harmonic oscillator is always greater than in the corresponding case of the free particle. 

\begin{figure}[t]
	\centering
	\includegraphics[width=0.99\linewidth]{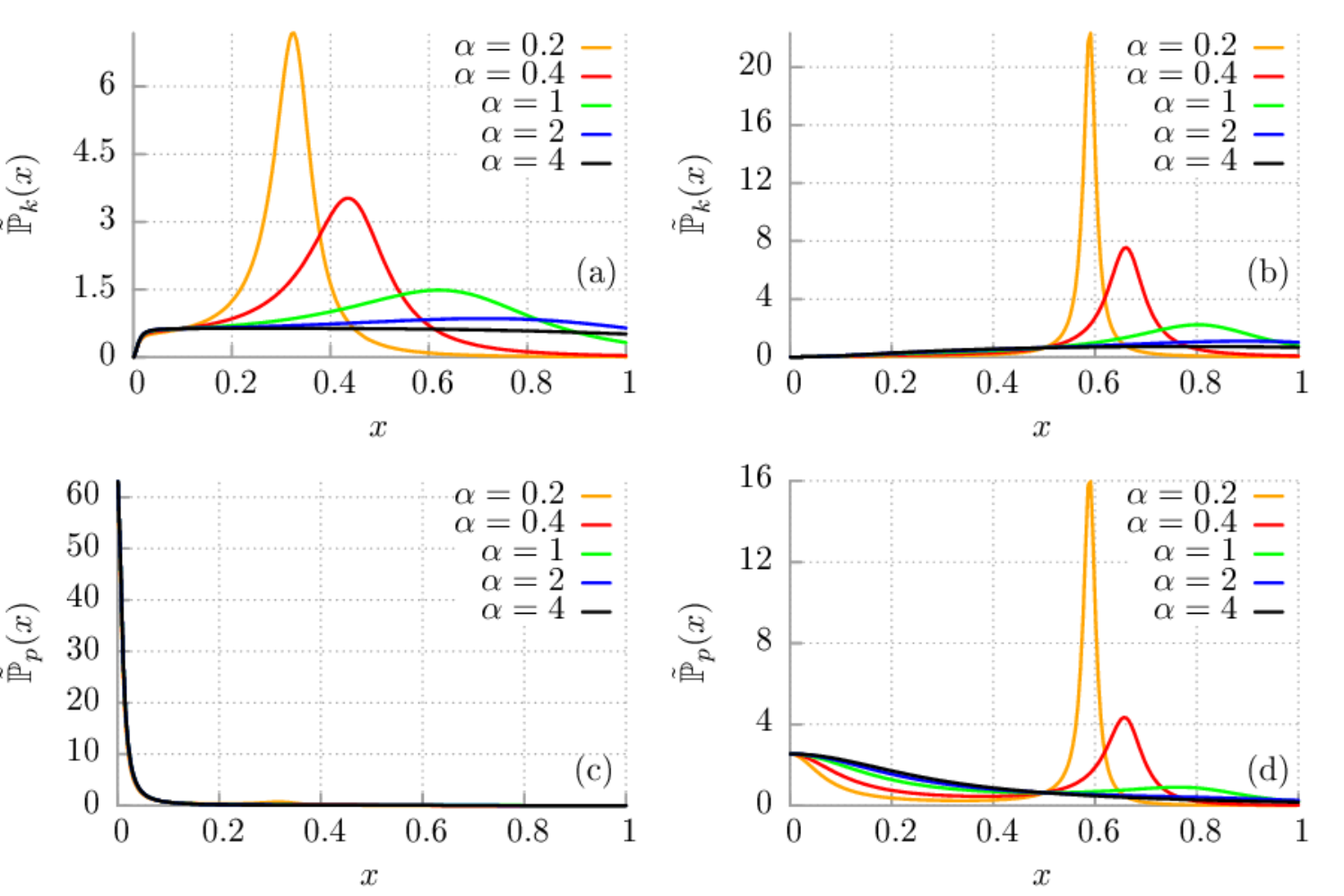}
		\caption{Drude model of dissipation. The probability distribution 
	$\mathbb{\tilde{P}}_k(x) = (1/\tau_v) \mathbb{P}_k(x/\tau_v)$ corresponding to the mean kinetic energy of the quantum harmonic oscillator is depicted  for selected  values of the parameter $\alpha=\tau_v/\tau_c$ with fixed $\tau_v=M/\gamma_0$ and    $\tilde{\omega}_0 = 0.1$ [panel (a)] and $\tilde{\omega}_0 = 0.5$ [panel (b)].  Panel (c) and (d): The probability density $\mathbb{\tilde{P}}_p(x)$ corresponding to the  mean  potential energy of the quantum harmonic oscillator is shown for different  $\alpha$, fixed  $\tilde{\omega}_0 = 0.1$ and $\tilde{\omega}_0 = 0.5$, respectively.}
	\label{fig3}
\end{figure}

In panels (c) and (d) of Fig. 1 we analyse the dependence of the total averaged energy $\tilde{E}=\tilde{E}_k+\tilde{E}_p$ of the quantum oscillator versus the previously discussed parameters. It is instructive to observe in the panel (c) that when the memory time $\tau_c$ decreases (i.e. $\alpha$ increases) the total energy of the system increases to infinity. It means that the limiting case of vanishing memory is non-physical for quantum systems. Since the time scale $\tau_c$ can be viewed also as the leading correlation time of the quantum thermal fluctuations one would say in analogy to classical physics that there is no limit of white noise in the quantum realm. In other words it implies that quantum thermal fluctuations are always correlated. Qualitatively, the dependence of the kinetic, potential or total energy on temperature is robust with respect to changes of the model parameter values. For high enough temperature it always tends to the classical limit $k_B T/2$ while in the regime of low temperature it is higher than the corresponding classical value. Note that all curves are monotonically increasing functions of temperature which never intersect each other. Due to this fact for a qualitative analysis it is sufficient to study the oscillator energies corresponding to zero temperature limit $T = 0$.

Here, we mention two recent papers \cite{pra_massignian,quantum_lampo} where  similar problems are studied. There the variance of the position of the quantum Brownian particle is studied as a function of temperature and the system-thermostat coupling strength.  One of the main results of analysis performed there is the particle position squeezing as temperature decreases and the interaction strength increases. For our system we observe a similar effect (not depicted). The potential energy $E_p$ (the particle position variance) decreases for fixed temperature $T$ and growing of the coupling constant $\gamma_0$. It then translates to the fact that the probability distribution $\mathbb{P}_p(\omega)$ corresponding to the mean potential energy rapidly decays meaning that relatively only the oscillators of low frequency bring the contribution to the average potential energy. Under this assumption they have small kinetic energy and therefore can transfer only little amount of it to the system. Consequently, the variance of the particle position is limited. In contrast, for weak system-thermostat coupling oscillators of high frequency dominate the probability distribution for the potential energy (position variance). Then they are allowed to have much larger kinetic energy and may transfer much bigger portion of it to the system resulting in increase of the particle position variance. Therefore the theorem of quantum partition of energy turns out to be quite helpful in qualitative interpretation of the mentioned particle position squeezing effect. 

The case of zero temperature $T=0$ is analysed in Fig. 2 where the impact of the memory time $\tau_c$ as well as the eigenfrequency $\tilde{\omega}_0$ is shown. Now additionally we compare the two mentioned mechanisms of dissipation. Panel (a) of this figure shows that when the memory time $\tau_c$ decreases (i.e. $\alpha$ increases) the kinetic energy monotonically increases. The opposite effect is for the potential energy: it slowly decreases as the memory time is shorter.  One can note that kinetic energy for Drude model is greater than for the algebraic decay of $\gamma(t)$. For the potential energy it is opposite sequence: $E_p$ is greater for the algebraic decay of $\gamma$ than for the exponential one. Moreover, both the kinetic as well as potential energy grows as the eigenfrequency $\tilde{\omega}_0$ is increased. Finally, the influence of the coupling strength  $\gamma_0$ should be pointed out (not shown in figures). It seems to be rather obvious that if the coupling is stronger then more channels are open to transmit energy from environment to the central system $S$ and therefore its energy is greater.

\begin{figure}[t]
	\centering
	\includegraphics[width=0.55\linewidth]{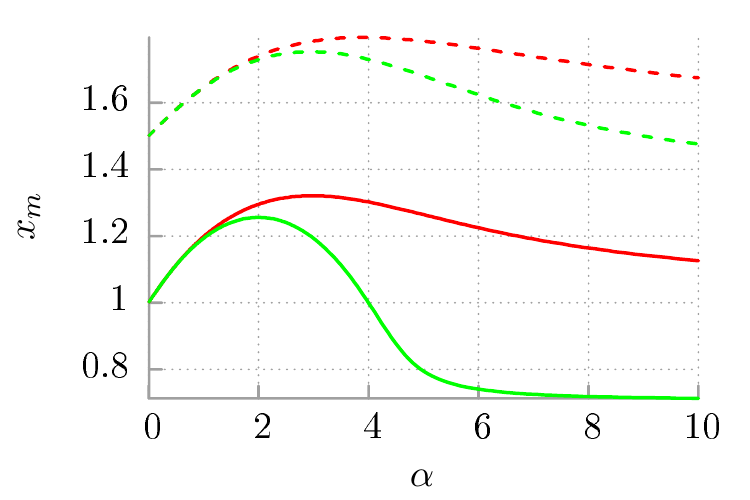}
	\caption{The dependence of the optimal frequency $x_m$ of the thermostat oscillators (at which the probability distribution is maximal)  on the parameter $\alpha$ for the Drude dissipation mechanism. The red and green colour corresponds to the kinetic $\tilde{E}_k$ and potential $\tilde{E}_p$ energy. The solid and dashed lines are for  $\tilde{\omega}_0 = 1$ and $\tilde{\omega}_0 = 1.5$, respectively.}
	\label{max}
\end{figure}

\subsection{Information provided in probability distributions}

In reach literature, formulas for the average kinetic and/or potential energy of the dissipative quantum harmonic oscillator appear in various context  in original papers and textbooks. We can mention several of them:  Eq. (83) in Ref. \cite{hakim}, Table 2 of Ref. \cite{ingold}, Eq. (4.14) in Ref. \cite{ford88} or Eq. (3.475) in Ref. \cite{breuer}. The expressions for the  mean kinetic and/or potential energy can also be obtained  directly or indirectly from different forms of fluctuation-dissipation relations \cite{call,kub,ford,zubarev} which are derived in the framework of the linear response theory which relates relaxation of a weakly perturbed system to the spontaneous fluctuations in thermal equilibrium, see e.g. Eq. (6.85) and (6.87) in Ref. \cite{weis} and Eq. (3.498) and (3.499) in Ref. \cite{breuer}. 
Therefore although the calculation of both kinetic and potential energy for a dissipative quantum oscillator has been done, the interpretation of these results as a quantum counterpart of the equipartition theorem expressed by the probability distributions $\mathbb{P}_k(x)$ or $\mathbb{P}_p(x)$ represents, to the best of our knowledge, an original point of view which may  help to improve the general understanding of the physics of dissipative quantum systems.

In Fig. \ref{fig3} we depict the dimensionless probability distribution $\tilde{\mathbb{P}}_k(x)$ and $\tilde{\mathbb{P}}_p(x)$ for the Drude dissipation mechanism and selected values of the model parameters.   The general observation is that the thermostat oscillators contribute to the energetics of the central system in a noticeably different way. In panel (a) and (b) we present the probability distribution $\tilde{\mathbb{P}}_k(x)$ corresponding to the kinetic energy $\tilde{E}_k$ of the oscillator. The reader can observe that if the memory time $\tau_c$ is large (i.e. $\alpha$ is small) then the probability distribution is peaked around some optimal frequency $x_m$ which brings the greatest contribution to the energy of system. On the other hand, if the memory time is getting smaller then the probability distribution is progressively flattened. The influence of the oscillator eigenfrequency $\tilde{\omega}_0$ is depicted in panel (b). We note that an increase of this parameter causes shifting of the density towards larger frequencies $x$. However, in each case the overall shape is conserved. A radically different behaviour is observed for the distribution $\tilde{\mathbb{P}_p}(x)$ corresponding to the potential energy $\tilde{E}_p$ of the oscillator. We illustrate it in the panel (c) and (d) of the same figure. In particular, we note that when the eigenfrequency $\tilde{\omega}_0$ of the system is small this distribution is robust with respect to changes of the memory time $\alpha$, c.f. panel (c). Then it is a rapidly decreasing function of the frequency which means that only thermostat oscillators of very small frequency contribute significantly to the potential energy of the system. It is distinctly different than it was in the case of the probability distribution for the kinetic energy. The situation changes for larger values of the oscillator eigenfrequency $\tilde{\omega}_0$, c.f. panel (d). Then for the long enough memory time (small $\alpha$) these two densities start to resemble each other. It means that both distributions are peaked and only thermostat oscillators taken from a very narrow interval of frequencies contribute to the corresponding energy of the system.

In Fig. \ref{max} we present the dependence of the optimal frequency $x_m$ of the thermostat oscillators (at $x_m$ the probability distribution has maximum) upon the parameter $\alpha$ for the Drude dissipation mechanism and selected values of the eigenfrequency $\tilde{\omega}_0$. For fixed values of $\tau_v =M/\gamma_0$ and $\omega_0$  the function  $x_m(\alpha)$ displays a non-monotonic character for the kinetic as well as the potential energy. It  means that there is a singled out value of  $\alpha$ ( i.e.  the memory time $\tau_c$) for which $x_m$ is maximal. We should stress that $x_m$ for the kinetic energy is greater than $x_m$ for the potential energy.   We can also conclude that for  large $\alpha$ oscillators of relatively lower frequency dominate in the contribution to both kinetic and potential energy.  In this panel we also depict the impact of the eigenfrequency $\tilde{\omega}_0$ on this characteristic. An increase of the latter parameter causes shifting of the curve towards larger values of $x_m$, however, the overall shape of the functional dependence remains unchanged.

\begin{figure}[t]
	\centering
	\includegraphics[width=0.99\linewidth]{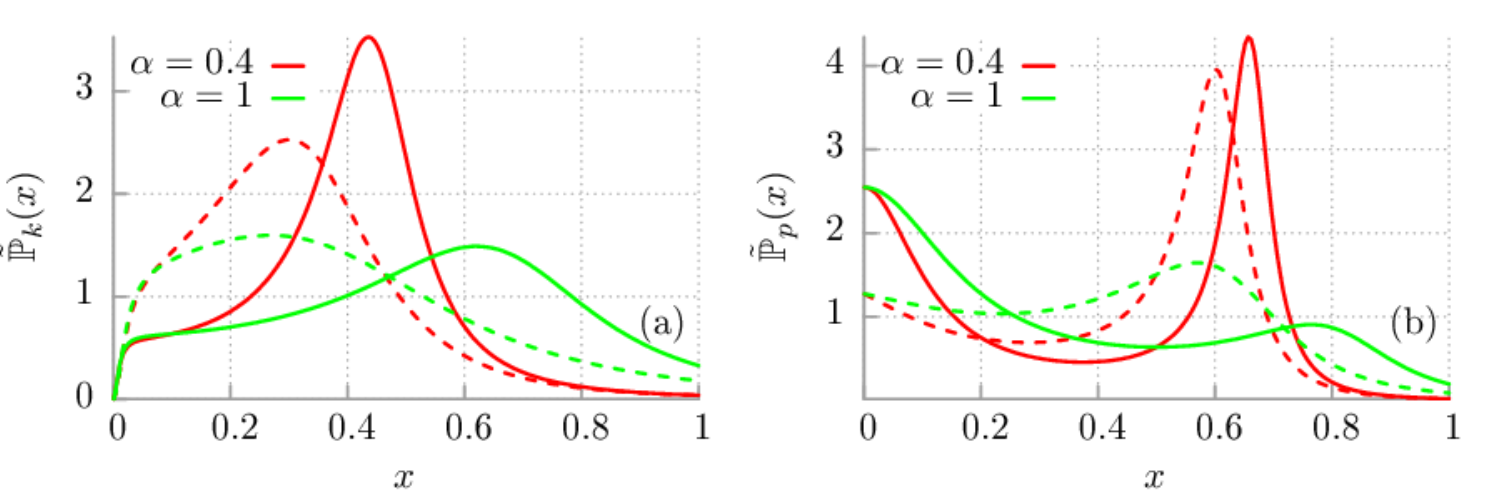}
		\caption{Comparison of the impact of Drude (solid lines) and algebraic (dashed lines) decay of the memory function. Panel (a): The probability distribution corresponding to the mean kinetic energy of the quantum harmonic oscillator is depicted for two  values of  $\alpha$ and for the frequency $\tilde{\omega}_0 = 0.1$.  Panel (b): The probability density $\mathbb{\tilde{P}}_p(x)$ corresponding to the  mean  potential energy of the quantum harmonic oscillator is shown for different magnitudes of the parameter $\alpha$  and $\tilde{\omega}_0 = 0.5$.}
	\label{fig4}
\end{figure}

Last but not least, in Fig. \ref{fig4} we compare the probability distributions $\tilde{\mathbb{P}}_k(x)$ and $\tilde{\mathbb{P}}_k(x)$ for the both considered dissipation mechanisms, i.e. exponential (Drude) and algebraic. The general remark is that the shape of the distributions $\tilde{\mathbb{P}}_k(x)$ and $\tilde{\mathbb{P}}_p(x)$ are qualitatively similar for the exponential and algebraic memory kernel. The difference is only visible in quantitative way. The important thing to note is that regardless of the values of the oscillator eigenfrequency $\tilde{\omega}_0$ the optimal frequencies which brings the most pronounced contribution to the kinetic as well as potential energy in the Drude model are higher than for the corresponding ones in the case of algebraic dissipation.

\section{Discussion}

We analysed partition of energy of the dissipative quantum harmonic oscillator. 
Mean kinetic and potential energy of the system are mean  kinetic and potential energies of the thermostat per one degree of freedom, i.e., 
$E_k = \langle \mathcal E_k \rangle$  and $E_p = \langle \mathcal E_p \rangle$, where 
$\langle \mathcal E_k \rangle$ and $ \langle \mathcal E_p \rangle$ are mean  kinetic and potential energies  per one degree of freedom of the thermostat which  consists of harmonic oscillators too. The symbol $\langle ...\rangle$ denotes two-fold averaging: 
 (i) over the Gibbs canonical state  for the thermostat  and (ii) over thermostat oscillators  frequencies $\omega$ which contribute to $E_k$ and $E_p$ according to the probability distribution $\mathbb{P}_k(\omega)$ and $\mathbb{P}_p(\omega)$, respectively.
The  relation for kinetic energy partition is similar to that  for classical
systems: The mean kinetic  energy of the oscillator equals the mean kinetic energy of the thermostat degree of freedom.  Of course, for classical systems the mean value of kinetic energy is $k_BT/2$ and depends only on temperature of thermostat. In the quantum case, it depends on "everything" (system-thermostat coupling, memory time, temperature). 
 
 We considered two examples of the dissipation mechanism: the Drude model characterised by the exponentially decreasing function and the algebraic decay of the memory kernel. We compared them and conclude that in the case of Drude model the kinetic energy of the oscillator is greater than for the algebraic decay. On the other hand, the reversed scenario is observed for the potential energy where the algebraic decay dominates the Drude dissipation. Moreover, the probabilities distributions are similar in both cases and display only quantitative differences. For the Drude model the optimal frequencies of thermostat oscillators which bring the largest contribution to the kinetic and potential energy are slightly higher than for the algebraic decay.

We have to emphasize that the quantum system which is open but not dissipative, does not obey the relations (\ref{Ek}) and (\ref{Ep}) for the energy partition. What we need is the thermodynamic limit for thermostat. The system is open when it interacts with environment but if the  environment is a system of finite degrees of freedom  then the memory function and the correlation function of quantum noise are quasi-periodic functions of time and the thermodynamic equilibrium state cannot be reached. In the case of finite thermostat, the response  function $R(t)$ in Eq. (\ref{p(t)}) is quasi-periodic, all three terms contribute to $\langle p^2(t)\rangle$  and the 
limit $t \to \infty$ does not exist for $\langle p^2(t)\rangle$.  

 One more issue should be discussed. When the memory time $\tau_c$ tends to zero, then the average energy of the oscillator increases to infinity. On the other hand, when $\tau_c \to 0$, the spectral density $J(\omega)$ is constant (the ohmic dissipation), the memory function $\gamma(t) \to \delta(t)$,  the  integro-differential Langevin equation  becomes local in time (as for classical Markovian processes) and is similar to a classical Newton equation with noise. However, we should also consider the correlation function $C(t)$ of noise $\eta(t)$. From Eq. (\ref{corr-f}) it follows that when $J(\omega)$ is constant then 
\begin{eqnarray} \label{CJ}
 C(t)\propto   \int_0^{\infty} d \omega \, \frac{\hbar \omega}{2} \coth \left(\frac{\hbar \omega}{2k_B T}\right) \cos(\omega t).    
\end{eqnarray} 
We see that it does not tend to white noise as in the classical case. It is even worse: it diverges!   We refer the interested reader to Ref. \cite{gardiner} for a more detailed analysis and to Ref. \cite{ingold} for discussion on the ohmic dissipation and Markovian limit.   Another aspect of the short memory time limit has been discussed for dynamics of solitons in superfluids \cite{efimkin}. 
This formal limit and the corresponding Markovian approximation gives rise to the Abraham-Lorentz force   (i.e., a term proportional to the derivative of the soliton’s acceleration) which  results in breaking of  causality. Three above non-physical effects  lead to the conclusion that  the limiting case of  vanishing memory time is not allowable  for quantum systems.

By the paradigmatic example of a quantum harmonic oscillator we demonstrated the quantum counterpart of the energy equipartition theorem which holds for classical systems. It is conceptually simple yet powerful tool for analysis of quantum open systems. Therefore we hope that our work in near future will open a new avenues within the area of physics.
\section{Methods} 
\subsection{Solutions of the generalized quantum Langevin equation}
In this section we present details of derivation of expressions for the averaged kinetic and potential energies of the quantum oscillator, i.e. Eqs. (\ref{Ek}) and (\ref{Ep}). 
The integral kernel of the GQLE (\ref{GLE}) is of convolution type and applying the Laplace transform yields the algebraic form, 
\begin{equation}\label{xL} 
\hat x_L(z) = \hat{R}_L(z) x(0)  +\hat{Q}_L(z)p(0) + \hat{Q}_L(z) \hat{\eta}_L(z), 
\end{equation}  
where 
$\hat{R}_L(z)$ and $\hat{Q}_L(z)$  are defined in Eq. (\ref{RL}). 
The inverse Laplace transform of  Eq. (\ref{xL}) gives the solution 
\begin{equation} \label{x(t)}
x(t) = R(t) x(0) + Q(t)p(0) + \int_0^t Q(t -s) \eta(s) ds.  
\end{equation}
From the theory of Laplace transform it follows that $\lim_{z\to\infty} \hat{f}(z) = 0$ for any function $f(t)$ for which the Laplace transform exists. In particular, it is also true for the functions $f_1(t) = \dot R(t) = dR(t)/dt$ and 
$f_2(t) = \dot Q(t)$. Calculating their  Laplace transform, we obtain  the relations
\begin{equation} \label{R(0)}
R(0) = \lim_{z\to\infty} z \hat{R}_L(z) = 1, \quad \quad 
Q(0) = \lim_{z\to\infty} z \hat{Q}_L(z) =0,  
\end{equation} 
which of course should be satisfied because of (\ref{x(t)}) for $t=0$. 
To derive a solution for the momentum $p(t)$ we note that the Laplace transform of the 
velocity is $\hat{v}_L(z) =  z \hat{x}_L(z) -x(0)$ and for the momentum one gets 
$\hat{p}_L(z) =  M z \hat{x}_L(z) - M x(0)$. We insert $\hat{x}_L(z)$ from Eq. (\ref{xL}) and utilize the equality for the Laplace transform of derivative of the function $\dot{R}(t)$, i.e.,  $\hat{\dot{R}}_L(z) = z \hat{R}_L(z) - 1$. The result is 
\begin{equation}\label{pL} 
\hat p_L(z) = M \hat{R}_L(z) x(0)  +\hat{Q}_L(z)p(0) + \hat{Q}_L(z) \hat{\eta}_L(z). 
\end{equation}  
Its inverse Laplace transform yields the solution 
\begin{equation}\label{p(t)} 
p(t) =  R(t)p(0) + M \dot{R}(t) x(0) + \int_0^t  R(t-s) \eta(s) ds, 
\end{equation}
Applying the similar method as above Eq. (\ref{pL}), one can show that $\dot{R}(0) =0$.

\subsection{Fluctuation-dissipation relation} 

Quantum noise $\eta(t)$ defined by Eq. (\ref{force}) is a family of non-commuting operators whose commutators are $c$-numbers. Its mean value over the Gibbs canonical state is zero, $\langle \eta(t) \rangle \equiv \mbox{Tr}\left[\eta(t)\rho_T\right] = 0$ and the symmetrized correlation function $C(t_1, t_2) = (1/2) \langle \eta(t_1)\eta(t_2)+\eta(t_2)\eta(t_1)\rangle \equiv \symkowm{\eta(t_1)}{\eta(t_2)}$ 
depends on the time difference,  $C(t_1, t_2) = C(t_1-t_2)=C(\tau)$. For 
$\tau=t_1-t_2$ it takes the form: 
\begin{eqnarray} \label{corr-f}
 C(\tau) = \sum_i \frac{\hbar c_i^2}{2 m_i \omega_i} \coth \left(\frac{\hbar \omega_i}{2k_B T} \right) \cos(\omega_i \tau) = \int_0^{\infty} d \omega \, \frac{\hbar \omega}{2} \coth \left(\frac{\hbar \omega}{2k_B T}\right) J(\omega)\cos \omega \tau,   
\end{eqnarray} 
where the spectral density $J(\omega)$ is defined in Eq. (\ref{spectral}). 
For an even function $f(t)$, we define  the  pair of Fourier cosine  transforms by the relations 
\begin{equation}\label{Four}
 f(t) = \int_0^\infty d\omega\, \hat{f}_F(\omega) \cos(\omega t), \qquad 
 \hat{f}_F(\omega) = (2/\pi) \int_0^\infty dt\,f(t) \cos(\omega t).
\end{equation}
We introduce the Fourier cosine transforms of the dissipation $\hat{\gamma}_F(\omega)$ and correlation  $\hat{C}_F(\omega)$ functions and compare them with  Eq. (\ref{g}). One notice that the following equality 
\begin{equation}\label{fludiss}
\hat{C}_F(\omega) = \frac{\hbar \omega}{2} \coth{\left( \frac{\hbar \omega}{2 k_B T} \right)} \hat{\gamma}_F(\omega)
\end{equation}
is satisfied. It is one of the form of the  fluctuation-dissipation theorem  \cite{call,kub,bialas,ford} (the extended discussion on this subject is also in Chapter 3 of Ref. \cite{zubarev}). It relates the memory kernel $\gamma(t)$ to the correlation function of the quantum thermostat noise $\eta(t)$ via its Fourier cosine transforms.  
On the other hand, $\hat{C}_F(\omega)$ is the Fourier transform of the correlation function $C(t)$ of the noise $\eta(t)$ and it is also called the power spectrum of noise.

\subsection{Potential  energy in an equilibrium state}
We calculate averaged potential energy of the quantum harmonic oscillator in the long time limit $t \to \infty$ when a thermal equilibrium state is reached. From Eq.  (\ref{x(t)})  we can obtain  the symmetrized position-position  correlation function $\symkowm{x(t)}{x(s)}$. 
For enough long times, i.e. much longer than the characteristic time scales  $\tau_v = M/\gamma_0$, $\tau_c$ and  $1/\omega_0$ only the last term of (\ref{x(t)}) contributes and then
\begin{equation} \label{xx}
\symkowm{x(t)}{x(s)} = \int_0^{t} dt_1 \int_0^{s} dt_2 \; Q(t-t_1) Q(s-t_2) \symkowm{\eta(t_1)}{\eta(t_2)}. 
\end{equation}
Now, we express the correlation function $C(t_1-t_2)= \symkowm{\eta(t_1)}{\eta(t_2)}$ by its Fourier cosine transform to get
\begin{eqnarray} \label{xTxS}
\symkowm{x(t)}{x(s)} = \int_0^{\infty} d\omega \; \hat{C}_F(\omega) \int_0^{t} dt_1 \int_0^{s} dt_2 \; Q(t-t_1)  Q(s-t_2) \cos \left[ \omega \left(t_1-t_2 \right) \right]. 
\end{eqnarray}
In particular, for $t=s$, it is the second statistical moment of the position operator, 
\begin{equation} \label{xTxT}
\langle x^2(t)\rangle = \int_0^{\infty} d\omega \; \hat{C}_F(\omega) \int_0^{t} dt_1 \int_0^{t} dt_2 \; Q(t-t_1) Q(t-t_2) \cos \left[ \omega \left(t_1-t_2 \right)
\right].
\end{equation}
We introduce new integration variables $\tau=t-t_1$ and $u= t-t_2$ and  convert equation (\ref{xTxT}) into the form 
\begin{equation} \label{x2}
\langle x^2(t)\rangle = \int_0^{\infty} d\omega \; \hat{C}_F(\omega) \int_0^{t} d\tau \int_0^{t} du \; Q(\tau) Q(u) \cos \left[ \omega \left(\tau-u \right) \right]. 
\end{equation}
We perform the limit  $t\to\infty$ and obtain the expression for the averaged potential  energy in the equilibrium state, namely,   
\begin{equation} \label{Epot}
E_p = \lim_{t \to \infty} \frac{1}{2} M\omega_0^2 \langle x^2(t)\rangle =
\frac{1}{2} M\omega^2_0 \int_0^{\infty} d\omega \; \hat{C}_F(\omega) I_p(\omega), 
\end{equation}
where 
\begin{equation} \label{Ip}
I_p(\omega) = \int_0^{\infty} d\tau \int_0^{\infty} du \; Q(\tau) Q(u) \cos \left[ \omega \left(\tau-u \right) \right]= \hat{Q}_L(i\omega) \hat{Q}_L(-i\omega)
\end{equation}
is the product of a Laplace transform of the response function $Q(t)$. To obtain the right hand side of this equation, we have exploited  relationship between the trigonometric functions and the complex exponential functions (the Euler's formula), and 
used the definition (\ref{fL}) for the Laplace transform. 

The next step is  use  the fluctuation-dissipation relation (\ref{fludiss}) to express the noise correlation spectrum $\hat{C}_F(\omega)$ by the dissipation spectrum $\hat{\gamma}_F(\omega)$. If we insert it to Eq. (\ref{Epot}) it becomes   
\begin{equation}\label{Epp}
E_p = \langle \mathcal{E}_p \rangle = \int_0^{\infty} d\omega \; \mathcal{E}_p(\omega)\mathbb{P}_p(\omega),  
\end{equation}
where  $\mathcal{E}_p(\omega)$ defined in Eq. (\ref{Ekp}) 
is thermal potential  energy per one degree of freedom of thermostat. The function 
 $\mathbb{P}_p(\omega)$ is given by  
\begin{eqnarray}\label{P_p}
\mathbb{P}_p(\omega) = M \omega^2_0  \hat{\gamma}_F(\omega)  
\hat{Q}_L(i\omega) \hat{Q}_L(-i\omega)  = 
\frac{i M \omega_0^2}{\pi \omega}  \left[\hat{Q}_L(i\omega) - \hat{Q}_L(-i\omega) \right]. 
\end{eqnarray}
The right hand side of this equations is obtained in the following way: In the left hand side, we express the Fourier cosine transform (\ref{Four})  by the Laplace transforms (\ref{fL}) for the function  $\hat{\gamma}_F(\omega) = 
(1/\pi) \, [\hat{\gamma}_L(i\omega) + \hat{\gamma}_L(-i\omega)]$. Next, we use the definition of $\hat Q_L(i\omega)$ and $\hat Q_L(-i\omega)$ in Eq. (\ref{RL}) and finally, after some algebra, we arrive to the result in (\ref{P_p}). 
In Appendix we show that $\mathbb{P}_p(\omega)$ fulfils all conditions to be a probability measure of some random variable. 

\subsection{Kinetic energy in an equilibrium state}

We proceed in the same way as in the previous subsection: by use of (\ref{p(t)}) we construct the symmetrized momentum-momentum correlation function $(1/2)\langle p(t)p(s) +p(s)p(t)\rangle$, exploit  the fluctuation-dissipation relation (\ref{fludiss}), take $t=s$ and perform the limit $t\to\infty$. The final result for the mean kinetic energy in a thermal equilibrium state is 
\begin{equation} \label{Ekin}
E_k = \lim_{t \to \infty} \frac{1}{2M} \langle p^2(t)\rangle =
\frac{1}{2M} \int_0^{\infty} d\omega \; \hat{C}_F(\omega) I_k(\omega), 
\end{equation}
where 
\begin{equation} \label{Ik}
I_k(\omega) = \int_0^{\infty} d\tau \int_0^{\infty} du \; R(\tau) R(u) \cos \left[ \omega \left(\tau-u \right) \right]= \hat{R}_L(i\omega) \hat{R}_L(-i\omega)
\end{equation}
is the product of a Laplace transform of the response function $R(t)$. In this equation, we convert the left side to the right side in a similar way as in Eq. (\ref{Ip}).  Now,  we again  use the  relation (\ref{fludiss}) to express  $\hat{C}_F(\omega)$ by the dissipation spectrum $\hat{\gamma}_F(\omega)$. Then (\ref{Ekin}) becomes 
\begin{equation}\label{Ekk}
E_k = \langle \mathcal{E}_k \rangle = \int_0^{\infty} d\omega \; \mathcal{E}_k(\omega)\mathbb{P}_k(\omega),  
\end{equation}
where  $\mathcal{E}_k(\omega)$ is  thermal kinetic energy per one degree of freedom of the thermostat (see Eq. (\ref{Ekp}))  and   
\begin{eqnarray}\label{P_k}
\mathbb{P}_k(\omega) = \frac{1}{M}  \hat{\gamma}_F(\omega)  
\hat{R}_L(i\omega) \hat{R}_L(-i\omega)   
= \frac{1}{\pi} \left[\hat{R}_L(i\omega) + \hat{R}_L(-i\omega) \right]. 
\end{eqnarray}
We convert the left side to the right side of this equation in a similar way as   Eq. (\ref{P_p}). 
In Appendix we prove that this function fulfils all conditions to be classified as a probability distribution of some random variable.  

\section{Appendix}

The functions $\mathbb{P}_p(\omega)$ defined by Eq. (\ref{P_p}) and $\mathbb{P}_k(\omega)$ defined by Eq. (\ref{P_k}) are both  probability densities on a  positive half-line of real numbers, i.e., they fulfil two conditions: 
\renewcommand{\theenumi}{\Alph{enumi}}
\begin{enumerate}
	\item non-negativity, $\mathbb{P}_{p,k}(\omega) \ge 0$, 
	\item normalization, $\int_0^{\infty} d\omega \; {\mathbb P}_{p,k}(\omega) = 1.$
\end{enumerate}

We can prove the non-negativity in the following way. In Eq. (\ref{P_p}) and  Eq. (\ref{P_k}) we use the definitions of $\hat Q_L(\pm i\omega)$ and 
$\hat R_L(\pm i\omega)$ in Eq. (\ref{RL}). For  $\hat{\gamma}_L(\pm i\omega)$ in these expressions we apply the relation $\hat{\gamma}_L(\pm i\omega) = A(\omega) \mp i B(\omega)$ with
\begin{eqnarray}  \label{Aom}
A(\omega) =  \int_0^{\infty} dt \; \gamma(t) \cos{(\omega t)},  \quad \quad 
B(\omega) = \int_0^{\infty} dt \; \gamma(t) \sin{(\omega t)}.
\end{eqnarray}
Then Eq. (\ref{P_p}) and  Eq. (\ref{P_k}) take the form
\begin{equation}\label{Pop}
\mathbb{P}_p(\omega) = \frac{2 M}{\pi} \frac{\omega_0^2 A{\left (\omega \right )}}{\omega^{2} A^{2}{\left (\omega \right )} + \left[M(\omega_0^{2}- \omega^{2}) + \omega B{\left (\omega \right )}\right]^{2}}, \quad 
\mathbb{P}_k(\omega) = \frac{2 M}{\pi} \frac{\omega^{2} A{\left (\omega \right )}}{\omega^{2} A^{2}{\left (\omega \right )} + \left[M(\omega_0^{2}- \omega^{2}) + \omega B{\left (\omega \right )}\right]^{2}}. 
\end{equation}
The denominator in (\ref{Pop}) is always positive and it is sufficient to show that the numerator $A(\omega) \ge 0$. 
From Eqs. (\ref{g}),  (\ref{Aom}) and (\ref{Four})  we deduce  that $A(\omega)=  (\pi/2) J(\omega)$. From Eq. (\ref{spectral}) it follows that  $J(\omega) \ge 0$ and the same holds true in the thermodynamic limit when $J(\omega)$ becomes a (piecewise) continuous function. Therefore $\mathbb{P}_{p,k}(\omega) \ge 0$.

The proof of the normalization condition is easier to perform for the distribution 
 $\mathbb{P}_k(\omega)$. From  Eq. (\ref{P})  one can obtain its equivalent form 
\begin{equation} \label{PkC}
 \mathbb{P}_k(\omega) = \frac{2}{\pi} \int_0^{\infty} dt\; R(t) \cos(\omega t)  
 = \hat{R}_C(\omega)  
\end{equation}
which is  a Fourier cosine  transform of the response function $R(t)$.   In turn, its inverse Fourier  transform reads 
\begin{equation} \label{R(t)}
 R(t) = \int_0^{\infty} d\omega \;  \hat{R}_C(\omega) \cos(\omega t). 
\end{equation}
From Eq. (\ref{R(0)}) it follows that $R(0) = 1$ and for $t=0$,   Eq. (\ref{R(t)}) reduces to  
\begin{equation}
R(0) = \int_0^{\infty} d\omega\; \hat{R}_C(\omega) = \int_0^{\infty} d\omega\; \mathbb{P}_k(\omega) = 1.
\end{equation}
 So, we proved the normalization of $ \mathbb{P}_k(\omega)$. Now, we prove it for 
 $\mathbb{P}_p(\omega)$. From Eq. (\ref{P_p}) one can obtain the 
representation of  $\mathbb{P}_p(\omega)$ in the form 
\begin{equation}\label{Psin}
\mathbb{P}_p(\omega) =  \frac{2 M\omega_0^2}{\pi} 
\int_0^{+\infty} dt \, Q(t) \, \frac{\sin(\omega t)}{\omega}.   
\end{equation}
By analogy  to Eq. (\ref{PkC}), we want to find such a function $V(t)$ that 
\begin{equation} \label{VQ}
 \int_0^{+\infty}  dt \, V(t) \cos(\omega t)  =  \int_0^{+\infty}  dt \, Q(t) \frac{\sin(\omega t)}{\omega}.  
\end{equation}
The first integral can be rewritten as 
\begin{equation}
 \int_0^{+\infty}  dt \, V(t) \frac{d}{dt} \left[ \frac{\sin(\omega t)}{\omega}\right]   =  - \int_0^{+\infty}  dt \, \frac{d V(t)}{dt} \;   \frac{\sin(\omega t)}{\omega}. 
\end{equation}
It  is true under conditions that $V(0)$ is bounded and $\lim_{t\to\infty} V(t)=0$. 
Then 
\begin{equation}
- \frac{d V(t)}{dt} = Q(t)  \quad \Rightarrow  \quad  V(t) = \int_t^{+\infty} Q(\tau) d \tau
\end{equation}
and it fulfils both conditions. In particular, $V(0) = \hat{Q}_L(0) = 1/M\omega_0^2$. 
From (\ref{Psin}) and (\ref{VQ}) one gets 
\begin{equation}\label{Pnor}
\mathbb{P}_p(\omega) =  \frac{2 M\omega_0^2}{\pi} 
\int_0^{+\infty} dt \, V(t) \cos(\omega t)   \quad \Rightarrow \quad 
V(t)  =  \frac{1}{ M\omega_0^2} \int_0^{+\infty} d\omega  \, \mathbb{P}_p(\omega) \cos(\omega t). 
\end{equation}
For $t=0$, it reduces to the relation 
\begin{equation}\label{Pnorma}
 \int_0^{+\infty} d\omega  \, \mathbb{P}_p(\omega) = M\omega_0^2 V(0) = 1. 
\end{equation}
\section*{Acknowledgement}
The work was supported by the Grants NCN 2015/19/B/ST2/02856 (P. B. and J. {\L}) and NCN 2017/26/D/ST2/00543 (J.S.) as well as the German Academic Exchange Service (DAAD) via scholarship in the program Research Stays for University Academics and Scientists (J.S.).
\section*{Author Contributions}
All authors carried  out  calculations, contributed to the discussion, analysis of the results and the writing of the manuscript.

\section*{Competing financial interests}
The authors declare no competing interests.


\begin{thebibliography}{99}
\bibitem{weis} Weiss, U. \textit{Quantum Dissipative Systems} (World Scientific, Singapore, 2008).
\bibitem{zurek} Zurek, W. H. Decoherence, einselection, and the quantum origins of the classical. \textit{Rev. Mod. Phys.} \textbf{75}, 715 (2003).
\bibitem{schlosshauer} Schlosshauer, M. Decoherence, the measurement problem, and interpretations of quantum mechanics. \textit{Rev. Mod. Phys.} \textbf{76}, 1267 (2005).
\bibitem{korbicz} Tuziemski, J. and Korbicz, J. K. Dynamical objectivity in quantum Brownian motion. \textit{EPL} \textbf{112}, 40008 (2015).
\bibitem{boy} Boyanovsky, D. and Jasnow, D. Heisenberg-Langevin versus quantum master equation. \textit{Phys. Rev. A} \textbf{96}, 062108 (2017).
\bibitem{smirne} Ferialdi, L. and Smirne, A. Momentum coupling in non-Markovian quantum Brownian motion. \textit{Phys. Rev. A} \textbf{96}, 012109 (2017).
\bibitem{carlesso} Carlesso, M. and Bassi, A. Adjoint master equation for quantum Brownian motion. \textit{Phys. Rev. A} \textbf{95}, 052119 (2017).  
\bibitem{china} Shen, H. Z., Su, S. L., Zhou, Y. H. and Yi, X. X. Non-Markovian quantum Brownian motion in one dimension in electric fields. \textit{Phys. Rev. A} \textbf{97}, 042121 (2018). 
\bibitem{lampo} Lim, S. H., Wehr, J.,  Lampo, A. and Lewenstein, M. On the Small Mass Limit of Quantum Brownian Motion with Inhomogeneous Damping and Diffusion. \textit{J. Stat. Phys.} \textbf{170}, 351 (2018).
\bibitem{huang} Huang, K. \textit{Statistical mechanics} (Wiley, New York, 1987).
\bibitem{terlecki} Terletski\'i, Y. P. \textit{Statistical Physics} (North-Holland: Amsterdam, The Netherlands, 1971). 
\bibitem{boltzmann} Boltzmann, L. \"Uber die Natur der Gasmolek\"ule. \textit{Wiener Berichte} \textbf{74}, 553 (1876).
\bibitem{arxiv} Bialas, P., Spiechowicz, J. and {\L}uczka, J. Quantum law for equipartition of energy. arXiv preprint at 	arXiv:1805.04012 (2018).
\bibitem{maga} Magalinskij, V. B. Dynamical Model in the Theory of Brownian Motion. \textit{J. Exp. Theor. Phys.} \textbf{36}, 1942 (1959).
\bibitem{ulersma} Ullersma, P. An exactly solvable model for Brownian motion: IV. Susceptibility and Nyquist's theorem. \textit{Physica} \textbf{32}, 27 (1966).
\bibitem{caldeira} Caldeira, A. and Leggett A. J. Path integral approach to quantum Brownian motion. \textit{Physica A} \textbf{121}, 587 (1983). 
\bibitem{ingold} Grabert, H., Schramm, P. and Ingold, G. L. Quantum Brownian Motion: The Functional Integral Approach. \textit{Phys. Rep.} \textbf{168}, 115 (1988).
\bibitem{hu} Hu, B. L, Paz, P. J. and Zhang, Y. Quantum Brownian motion in a general environment: Exact master equation with nonlocal dissipation and colored noise. \textit{Phys. Rev. D} \textbf{45}, 2843 (1992).
\bibitem{hujcen} Nieuwenhuizen, Th. M. and Allahverdyan, A. E. Statistical thermodynamics of quantum Brownian motion: Construction of perpetuum mobile of the second kind. \textit{Phys. Rev. E} \textbf{66}, 036102 (2002).
\bibitem{chaos} {\L}uczka, J. Non-Markovian stochastic processes: Colored noise. \textit{Chaos} \textbf{15}, 026107 (2005).
\bibitem{ph} H\"anggi, P. and Ingold, G. L. Fundamental Aspects of Quantum Brownian Motion. \textit{Chaos} \textbf{15}, 026105 (2005).
\bibitem{bialas} Bialas, P. and {\L}uczka, J. Kinetic energy of a free quantum Brownian particle. \textit{Entropy} \textbf{20}, 123 (2018).
\bibitem{feynman} Feynman, R. P. \textit{Statistical Mechanics} (Westview Press: Reading, PA, USA, 1972). 
\bibitem{quantum_lampo} Lampo, A., Lim, S. H., Garcia-March, M. A. and Lewenstein M. Bose polaron as an instance of quantum Brownian motion. \textit{Quantum} \textbf{1}, 30 (2017).
\bibitem{pra_massignian} Massignian, P., Lampo, A., Wehr, J. and Lewenstein M. Quantum Brownian motion with inhomogeneous damping and diffusion. \textit{Phys. Rev. A} \textbf{91}, 033627 (2015).

\bibitem{hakim} Hakim, V. and Ambegaokar, V. Quantum theory of a free particle interacting with a linearly dissipative environment.\textit{Phys. Rev. A} \textbf{32}, 423 (1985).
\bibitem{ford88} Ford, G. W., Lewis J. T. and O'Connell, R. F. Quantum oscillator in a blackbody radiation field II. Direct calculation of the energy using the fluctuation-dissipation theorem. \textit{Ann. Phys. (N.Y.)} \textbf{185}, 270 (1988).
\bibitem{breuer} Breuer, H. P. and Petruccione, F. \textit{The theory of open quantum systems} (New York, Oxford University Press, 2002).
\bibitem{call} Callen, H. B. and Welton, T. A. Irreversibility and Generalized Noise. \textit{Phys. Rev. 83}, \textbf{34} (1951).
\bibitem{kub} Kubo, R. The fluctuation-dissipation theorem. \textit{Rep. Prog. Phys.} \textbf{29}, 255 (1966). 
\bibitem{ford} Ford, G. W. The fluctuation–dissipation theorem. \textit{Contemporary Physics} \textbf{58}, 244 (2017). 
\bibitem{zubarev} Zubarev, D. N. \textit{Nonequilibrium statistical thermodynamics} (New York, Consultants Bureau, 1974).
\bibitem{gardiner}  C. W. Gardiner and P. Zoller, \textit{Quantum Noise} (Berlin, Springer-Verlag, 2004). 
\bibitem{efimkin} Efimkin, D., Hofmann, J. and Galitski V. Non-Markovian Quantum Friction of Bright Solitons in Superfluids. \textit{Phys. Rev. Lett.} \textbf{116}, 225301 (2016).

\end{thebibliography}
\end{document}